\documentclass{article}

\usepackage{arxiv}

\usepackage[utf8]{inputenc} % allow utf-8 input
\usepackage[T1]{fontenc}    % use 8-bit T1 fonts
\usepackage{hyperref}       % hyperlinks
\usepackage{url}            % simple URL typesetting
\usepackage{booktabs}       % professional-quality tables
\usepackage{amsfonts}       % blackboard math symbols
\usepackage{nicefrac}       % compact symbols for 1/2, etc.
\usepackage{microtype}      % microtypography
\usepackage{lipsum}		% Can be removed after putting your text content
\usepackage{graphicx}
\usepackage[numbers]{natbib}
\usepackage{doi}
\usepackage{float}
\usepackage{amsmath} 

\usepackage{pifont}

\usepackage{listings}
\usepackage{xcolor}
\definecolor{codegreen}{rgb}{0,0.6,0}
\definecolor{codegray}{rgb}{0.5,0.5,0.5}
\definecolor{codepurple}{rgb}{0.58,0,0.82}
\definecolor{backcolour}{rgb}{1,1,1}
\lstdefinestyle{mystyle}{
    backgroundcolor=\color{backcolour},   
    commentstyle=\color{codegreen},
    keywordstyle=\color{magenta},
    numberstyle=\tiny\color{codegray},
    stringstyle=\color{codepurple},
    basicstyle=\ttfamily\footnotesize,
    breakatwhitespace=false,         
    breaklines=true,                 
    captionpos=t, % put captions to the top               
    keepspaces=true,                 
    numbers=none, % none, or put line numbers to left             
    numbersep=5pt,                  
    showspaces=false,                
    showstringspaces=false,
    showtabs=false,                  
    tabsize=2
}
\makeatletter
\pretocmd\lst@makecaption{\noindent{\rule{\linewidth}{1pt}}}{}{}
\makeatother

\usepackage{rotating}

\title{Bridging the Gap Between PHE and FHE: A Performance and Trade-off Analysis of The Somewhat Homomorphic BGN Cryptosystem}

\author{ \href{https://orcid.org/0000-0002-0345-0088}{\includegraphics[scale=0.06]{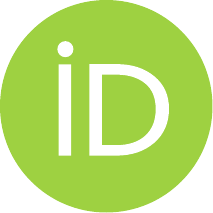}\hspace{1mm}Sefik Serengil} \\
	Department of Engineering\\
	Neo4j\\
    London, SE1 0LH, UK \\
	\texttt{sefik.serengil@neo4j.com} \\
	\And
	\href{https://orcid.org/0000-0003-1250-5949}{\includegraphics[scale=0.06]{orcid.pdf}\hspace{1mm}Alper Ozpinar} \\
	Department of Management\\
	Ibn Haldun University\\
    Istanbul, 34480, Turkiye \\
	\texttt{alper.ozpinar@ihu.edu.tr} \\
}

\renewcommand{\shorttitle}{Bridging PHE and FHE: Performance Analysis of BGN}

\hypersetup{
    pdftitle={Bridging the Gap Between PHE and FHE: A Performance and Trade-off Analysis of The Somewhat Homomorphic BGN Cryptosystem},
    pdfsubject={cs.CR},
    pdfauthor={Sefik Serengil, Alper Ozpinar},
    pdfkeywords={Cryptography, Homomorphic Encryption, Cloud Benchmarks},
}

\AtBeginDocument{
  \setlength{\abovedisplayskip}{2\baselineskip}
  \setlength{\belowdisplayskip}{2\baselineskip}
  \setlength{\abovedisplayshortskip}{2\baselineskip}
  \setlength{\belowdisplayshortskip}{2\baselineskip}
}

\begin{document}
\maketitle

\begin{abstract}
Homomorphic encryption stands as a pivotal paradigm for privacy-preserving data analytics, yet practitioners are frequently forced into a polarizing choice between lightweight Partially Homomorphic Encryption (PHE) and computationally dominant Fully Homomorphic Encryption (FHE). The Boneh-Goh-Nissim (BGN) cryptosystem theoretically bridges this divide as a Somewhat Homomorphic Encryption (SWHE) scheme by supporting unlimited additions and exactly one ciphertext-ciphertext multiplication. However, despite its profound algebraic elegance, the practical adoption of BGN has remained stagnant due to the lack of accessible, production-ready software implementations. This paper introduces a rigorous comparative trade-off analysis of the BGN cryptosystem against both PHE and FHE paradigms, evaluated through its definitive open-source integration into the \texttt{lightphe} Python framework. Through this integration, a fully functional BGN cryptosystem can be initialized and deployed for homomorphic operations within just a few lines of Python code, drastically reducing the implementation barrier. We benchmark encrypted vector operations over a 128-dimensional space under strict 80-bit and 112-bit security compliance, cross-evaluating BGN against prominent PHE schemes—specifically Paillier, Damg{\aa}rd-Jurik, and Okamoto-Uchiyama—as well as the FHE-backed CKKS scheme via TenSEAL. Our empirical results expose a striking computation-communication-precision paradox across the cryptographic spectrum. We explicitly demonstrate that, computationally, BGN is heavily penalized by expensive bilinear pairing evaluations, making it significantly slower than both traditional PHE counterparts and modern FHE frameworks that leverage highly optimized C++ SIMD architectures. However, our integrated BGN framework successfully capitalizes on a profound communication and architectural advantage: it retains a microscopic public key footprint of just 3 to 6 KB—an improvement of up to five orders of magnitude over FHE. Crucially, we demonstrate that although BGN permits only a single ciphertext multiplication, the resulting product terms are projected onto the identical target algebraic plane, allowing them to be homomorphically aggregated without bounds. This structural harmony enables the seamless execution of complex multivariate algorithms such as linear regression inference, Cosine Similarity, and Squared Euclidean Distance metrics over fully encrypted payloads. Furthermore, we demonstrate that an optimized precision of 2 digits in BGN is entirely sufficient to match plaintext ranking baselines, effectively neutralizing the target-group discrete logarithm decryption bottleneck. By open-sourcing this pipeline, this work democratizes the BGN scheme, establishing \texttt{lightphe} as the foundational engine for future SWHE-driven decentralized and bandwidth-constrained architectures where usability and minimal communication footprints outweigh raw server-side computational latency.
\end{abstract}

\keywords{Cryptography, Homomorphic Encryption, Somewhat Homomorphic Encryption, BGN, Python, Privacy-Preserving Machine Learning, Privacy-Preserving Vector Search, Homomorphic Vector Embeddings, Federated Learning}

\section{Introduction}

The rapid proliferation of Privacy-Preserving Machine Learning (PPML), decentralized edge architectures, and collaborative frameworks like Federated Learning (FL) has fundamentally transformed modern data analytics. Moreover, with the exponential rise of Large Language Models (LLMs) and Retrieval-Augmented Generation (RAG) pipelines, organizations increasingly rely on Vector Databases to index and query high-dimensional data embeddings. However, outsourcing sensitive multi-dimensional vectors—such as clinical medical records, financial transactions, biometric profiles, or proprietary LLM vector embeddings—to untrusted cloud infrastructures introduces severe privacy, security, and compliance risks. Traditional cryptographic frameworks secure data at rest and in transit but fail when data must be processed, forcing systems to decrypt information in memory and creating a prime target for runtime exploits.

Homomorphic Encryption (HE) offers a definitive mathematical solution to this paradigm by enabling computation directly on encrypted payloads without requiring access to secret keys. In the context of Privacy-Preserving Vector Search and biometric retrieval, maintaining the privacy of Homomorphic Vector Embeddings during similarity scoring (e.g., Cosine Similarity or Distance evaluation) has become a primary bottleneck.

Despite its immense theoretical promise, the practical deployment of homomorphic encryption is frequently hindered by a stark architectural polarization within the cryptographic spectrum. On one end of this spectrum reside Partially Homomorphic Encryption (PHE) schemes \cite{phe}, which natively support only a single type of algebraic operation (either addition or multiplication) with an arbitrary execution depth. Conversely, Fully Homomorphic Encryption (FHE) \cite{fhe} frameworks accommodate arbitrary circuits comprising both additions and multiplications, but introduce an overwhelming operational tax known as noise management \cite{ckks} \cite{bgv}. Every ciphertext-ciphertext multiplication in FHE injects algebraic noise into the underlying data structure; preventing its corruption requires complex bootstrapping techniques or highly parameterized, deep slot allocations that inflate ciphertext structures. Consequently, deploying FHE pipelines requires massive public keys (frequently scaling from tens to hundreds of megabytes) and substantial communication bandwidth. This creates a severe computation-communication paradox, disqualifying FHE from being used in bandwidth-constrained environments, Internet of Things (IoT) nodes, and real-time edge computing architectures.

The Boneh-Goh-Nissim (BGN) cryptosystem, introduced as a Somewhat Homomorphic Encryption (SWHE) paradigm, theoretically bridges this widening chasm between the restrictive linearity of PHE and the prohibitive communication overhead of FHE \cite{bgn}. By leveraging the elegant mathematics of bilinear pairings over elliptic curves, BGN accommodates an arbitrary, unlimited depth of homomorphic additions alongside exactly one layer of ciphertext-ciphertext multiplication. Crucially, because all resulting product terms are projected onto the identical target algebraic plane via the pairing operation, these post-multiplication ciphertexts can be homomorphically aggregated and accumulated without bounds. This specific cross-term additive harmony provides the exact algebraic blueprint required to evaluate fundamental multivariate algorithms—such as linear regression inference, Cosine Similarity (dot product), and Squared Euclidean Distance metrics—over fully encrypted multi-dimensional payloads. Furthermore, BGN achieves this without the multi-megabyte key inflation or noise-management overhead characteristic of FHE, maintaining a microscopic public key footprint that mirrors the lightweight nature of traditional PHE schemes.

Yet, despite its profound algebraic elegance and clear utility for vector analytics, the practical adoption of the BGN cryptosystem has remained stagnant for over two decades. The root of this stagnation is not mathematical, but software-centric. The vast majority of cryptographic toolkits either ignore BGN entirely or relegate it to archaic, low-level C++ academic proof-of-concepts that require profound manual parameter tuning and lack integration with modern data science languages. Researchers and data engineers working in Python—the undisputed language of modern machine learning and analytics—have been completely locked out of deploying BGN in actual production pipelines. 

To systematically address this software barrier and democratize access to advanced homomorphic structures, this paper introduces the definitive open-source integration and empirical evaluation of the BGN cryptosystem within the \texttt{lightphe} Python framework \cite{lightphe}. Developed to serve the broader scientific and industrial communities, \texttt{lightphe} is distributed under the permissive MIT License and hosted as a public, open-source project on GitHub at \url{https://github.com/serengil/LightPHE}, thereby ensuring unrestricted academic reproducibility and cross-industry utility. The \texttt{lightphe} ecosystem functions as a comprehensive, abstract computational engine designed to provide uniform access to a wide array of homomorphic algorithms. By embedding BGN directly into this operator-overloaded framework, a fully functional SWHE environment can now be initialized, managed, and deployed for vector analytics within just a few lines of clean Python code. To contextualize the versatility of this framework, \texttt{lightphe} natively orchestrates an unparalleled, exhaustive suite of cryptographic paradigms, establishing a global benchmarking environment that includes:

\begin{itemize}
    \item \textbf{Multiplicative PHE Algorithms:} Classic unpadded RSA \cite{rsa} and standard ElGamal \cite{elgamal} implementations.
    \item \textbf{Additive \& Advanced ElGamal Variants:} Exponential ElGamal alongside Elliptic Curve ElGamal (EC-ElGamal) \cite{ecelgamal}, featuring full algebraic support for Short Weierstrass \cite{weierstrass}, Koblitz \cite{koblitz}, and Twisted Edwards \cite{edwards} curve coordinates to optimize scalar point multiplications.
    \item \textbf{Advanced Additive PHE Schemes:} The widely adopted Paillier cryptosystem \cite{paillier}, its arbitrary higher-power integer generalization via the Damg{\aa}rd-Jurik scheme \cite{damgard}, and Okamoto–Uchiyama \cite{okamoto}.
    \item \textbf{Specialized Historic \& Algebraic PHE Paradigms:} The Benaloh cryptosystem \cite{benaloh} (supporting dense plaintext blocks), the Naccache–Stern \cite{naccache} higher-radix framework, the Goldwasser–Micali \cite{goldwasser} probabilistic bit encryption scheme, and the Sander–Young–Yung (SYY) \cite{sander} homomorphic log-depth circuit evaluator.
    \item \textbf{Somewhat Homomorphic Encryption:} The newly integrated Boneh-Goh-Nissim (BGN) \cite{bgn} elliptic curve pairing-based engine.
\end{itemize}

As systematically codified in Table~\ref{tab:homomorphic_taxonomy}, the homomorphic capabilities across the comprehensive \texttt{lightphe} ecosystem exhibit distinct algebraic boundaries. While traditional frameworks strictly isolate additive or multiplicative properties, the newly integrated Boneh-Goh-Nissim cryptosystem uniquely accommodates an infinite sequence of additions intertwined with exactly one layer of homomorphic multiplication, thereby executing complex multivariate calculations over uniform pipelines.

\begin{table}[htbp]
\centering
\caption{Homomorphic Capability Taxonomy of Cryptosystems}
\label{tab:homomorphic_taxonomy}
\small
\begin{tabular}{lccccc}
\hline
\textbf{Algorithm} & \textbf{\begin{tabular}[c]{@{}c@{}}Multiplicatively\\ Homomorphic\end{tabular}} & \textbf{\begin{tabular}[c]{@{}c@{}}Additively\\ Homomorphic\end{tabular}} & \textbf{\begin{tabular}[c]{@{}c@{}}Scalar\\ Multiplication\end{tabular}} & \textbf{\begin{tabular}[c]{@{}c@{}}Bitwise-XOR\\ Homomorphic\end{tabular}} & \textbf{\begin{tabular}[c]{@{}c@{}}Bitwise-AND\\ Homomorphic\end{tabular}} \\ \hline
RSA & \checkmark & $\times$ & $\times$ & $\times$ & $\times$ \\
ElGamal & \checkmark & $\times$ & $\times$ & $\times$ & $\times$ \\
Exponential ElGamal & $\times$ & \checkmark & \checkmark & $\times$ & $\times$ \\
Elliptic Curve ElGamal & $\times$ & \checkmark & \checkmark & $\times$ & $\times$ \\
Paillier & $\times$ & \checkmark & \checkmark & $\times$ & $\times$ \\
Damg{\aa}rd-Jurik & $\times$ & \checkmark & \checkmark & $\times$ & $\times$ \\
Benaloh & $\times$ & \checkmark & \checkmark & $\times$ & $\times$ \\
Naccache-Stern & $\times$ & \checkmark & \checkmark & $\times$ & $\times$ \\
Okamoto-Uchiyama & $\times$ & \checkmark & \checkmark & $\times$ & $\times$ \\
Goldwasser-Micali & $\times$ & $\times$ & $\times$ & \checkmark & $\times$ \\
Sander-Young-Yung & $\times$ & $\times$ & $\times$ & $\times$ & \checkmark \\
\hline
\textbf{Boneh-Goh-Nissim} & \textbf{1} & \checkmark & \checkmark & $\times$ & $\times$ \\ \hline
\end{tabular}
\end{table}

In our baseline comparisons, we restricted the PHE spectrum to the Paillier, Damgård-Jurik, and Okamoto-Uchiyama cryptosystems. Other additively homomorphic schemes—such as Benaloh, Naccache-Stern, Exponential-ElGamal, and Elliptic Curve ElGamal (EC-ElGamal)—were deliberately excluded due to critical practical limitations. Specifically, schemes like Benaloh and Naccache-Stern incur substantial computational overhead for high-dimensional vector representations. Meanwhile, Exponential-ElGamal and EC-ElGamal require solving the Discrete Logarithm Problem (DLP) during the decryption phase; as vector dimensions and floating-point precision scale, resolving the plaintext value via Pollard's rho or brute-force search becomes computationally prohibitive for real-time applications. Conversely, Paillier, Damgård-Jurik, and Okamoto-Uchiyama offer direct algebraic decryption without DLP search bottlenecks, alongside seamless support for scalar multiplication required in encrypted-plain vector dot products.

The primary contribution of this work is a rigorous, three-way empirical benchmark that maps the exact trade-off boundaries between PHE (specifically benchmarking against Paillier, Damg{\aa}rd-Jurik, and Okamoto–Uchiyama), SWHE (via our BGN integration), and FHE (via TenSEAL's CKKS implementation \cite{tenseal}). We evaluate these paradigms over a 128-dimensional encrypted vector space under strict 80, 112 and 128-bit cryptographic security compliance, exploring the multi-dimensional trade-offs between computational throughput, public key size, and decimal precision. Our results expose a dramatic communication-precision trade-off: while TenSEAL achieves rapid SIMD packaging, it demands public keys up to 451 MB. In stark contrast, our BGN framework operates with a microscopic key size of just 3 to 6 KB—a reduction of five orders of magnitude—while fully maintaining numerical ranking fidelity at an optimized decimal precision of 2. By delivering this pipeline to the open-source community, this work removes the mathematical pairing barriers surrounding BGN and establishes \texttt{lightphe} as the definitive computational platform for future bandwidth-constrained, privacy-preserving analytical architectures.

\section{Mathematical Foundations of the BGN Cryptosystem}

The Boneh-Goh-Nissim (BGN) cryptosystem relies on the algebraic properties of bilinear pairings \cite{pairing_miller} evaluated over supersingular elliptic curves. Unlike traditional Partially Homomorphic Encryption (PHE) frameworks that operate entirely within a single algebraic group, BGN utilizes an evaluation pipeline that maps elements contextually from a source elliptic curve group to a multiplicative target field extension \cite{pairing_boneh}. This section formalizes the cryptographic primitives, key generation constraints, operational algebraic transformations, and decryption mechanics instantiated within the framework.

\subsection{Key Generation and Curve Constraints}
The setup phase mandates the selection of a composite modulus that governs the underlying algebraic structures. The execution sequence proceeds as follows:
\begin{enumerate}
    \item Select two large, distinct cryptographic odd primes $q_1$ and $q_2$ of equal bit-length, and compute the composite RSA modulus:
    \begin{equation}
        n = q_1 \times q_2
    \end{equation}
    \item Construct a large prime $p$ acting as the finite field characteristic such that:
    \begin{equation}
        p = n \times l - 1
    \end{equation}
    where $l \in \mathbb{Z}^+$ is a systematic multiple of 4, ensuring that $p \equiv 3 \pmod 4$. This precise congruence is an absolute mathematical prerequisite to guarantee the supersingularity of the chosen elliptic curve.
    \item Define a supersingular elliptic curve over the prime field $\mathbb{F}_p$ via the Weierstrass form:
    \begin{equation}
        E(\mathbb{F}_p): y^2 = x^3 + x
    \end{equation}
    where the curve parameters are strictly bound to $a=1$ and $b=0$. The total number of points on this curve is given by $\#E(\mathbb{F}_p) = p + 1 = n \times l$, meaning the curve order is a direct multiple of the composite modulus $n$.
    \item Locate a base generator point $G \in E(\mathbb{F}_p)$ of order $n$. To achieve this, a random point on the curve is scaled by the cofactor $l$. 
    \item Generate the blinding parameters by choosing a random integer $r \in [2, n-1]$ such that $\gcd(r, n) = 1$. The public blinding point $u$ is evaluated via scalar multiplication as $u = r \cdot G$. The final structural public key component $h$ is computed by scaling $u$ with the hidden prime factor $q_2$:
    \begin{equation}
        h = q_2 \cdot u = (q_2 \cdot r) \cdot G
    \end{equation}
    Due to this construction, the point $h$ possesses a strict order of $q_1$ within the group, since $n \cdot u = (q_1 \cdot q_2 \cdot r) \cdot G = \mathcal{O}$.
\end{enumerate}
The resulting \textbf{Public Key} is exported as $(E, G, n, h, l)$, while the corresponding \textbf{Private Key} is the prime pair $(q_1, q_2)$.

It is critical to distinguish the structural properties of the BGN cryptosystem from standard Elliptic Curve Cryptography (ECC). In mainstream ECC operating over prime fields (e.g., ECDSA or Ed25519), security relies entirely on the Elliptic Curve Discrete Logarithm Problem (ECDLP). This allows ECC to achieve equivalent security tiers with drastically smaller key footprints compared to factoring-based schemes; for instance, a 160-bit or 224-bit ECC key offers security parity with 1024-bit or 2048-bit RSA keys, respectively. 

Conversely, the BGN cryptosystem deviates from this lightweight key convention due to its foundational reliance on both the Subgroup Decision Problem and the factorization of the composite RSA modulus $n = q_1 \times q_2$. Because the underlying algebraic infrastructure utilizes a supersingular elliptic curve where the order is bound to a multiple of $n$, BGN does not inherit the compact key-size advantage typical of standard ECC. Instead, its key scaling and security parameters are strictly constrained by RSA factorization bounds. To guarantee robust 128-bit security compliance, the composite modulus $n$ must span at least 3072 bits, reflecting a key size paradigm that mirrors RSA variants rather than conventional ECC frameworks.

\subsection{Homomorphic Encryption and Layer-1 Addition ($\mathbb{G}_1$)}
Given a plaintext integer message $m \in \mathbb{Z}_n$, encryption requires a fresh, single-use random blinding factor $r \in [1, q_2 - 1]$. The ciphertext $C$ is constructed as a point on the elliptic curve group $\mathbb{G}_1 = E(\mathbb{F}_p)$ via:
\begin{equation}
    C = (m \cdot G) + (r \cdot h) \;\; \in \mathbb{G}_1
\end{equation}
Homomorphic addition of two independent Layer-1 ciphertexts, $C_1 = (m_1 \cdot G) + (r_1 \cdot h)$ and $C_2 = (m_2 \cdot G) + (r_2 \cdot h)$, is achieved directly via standard elliptic curve point addition:
\begin{equation}
    C_{\text{add}} = C_1 + C_2 = (m_1 + m_2) \cdot G + (r_1 + r_2) \cdot h
\end{equation}
Because $(r_1 + r_2)$ acts as a combined valid blinding factor, $C_{\text{add}}$ constitutes a structurally sound encryption of $(m_1 + m_2)$. Scalar multiplication of a ciphertext by a plaintext constant $k$ is executed via standard point scaling:
\begin{equation}
    C_{\text{scalar}} = k \cdot C_1 = (k \cdot m_1) \cdot G + (k \cdot r_1) \cdot h
\end{equation}
This confirms that linear evaluations can be executed to an arbitrary, infinite depth within the source curve group $\mathbb{G}_1$.

\subsection{Bilinear Pairings and Layer-2 Multiplication ($\mathbb{G}_T$)}
\begin{figure}[H]
    \centering
    \includegraphics[width=0.8\linewidth]{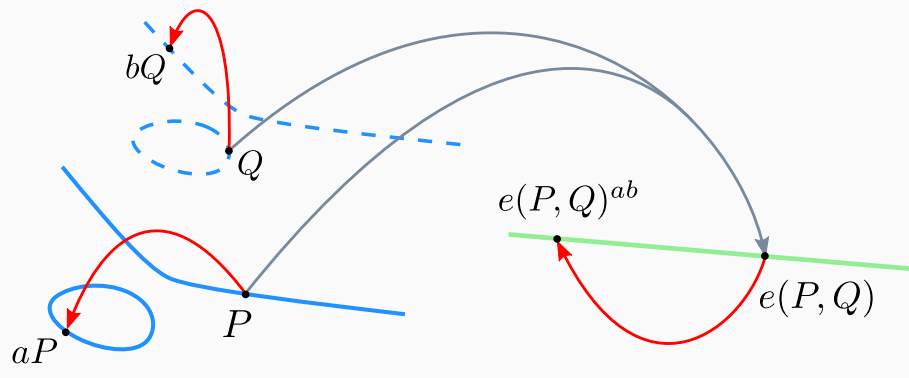}
    \caption{Elliptic Curve Pairings \cite{pairing_illustration}}
    \label{fig:pairings}
\end{figure}

The core breakthrough of BGN is its ability to perform a non-linear homomorphic multiplication between two independent ciphertexts. Figure \ref{fig:pairings} presents a high-level overview of how elliptic curve pairings are performed. This is governed by a modified Weil or Tate pairing acting as a non-degenerate bilinear map:
\begin{equation}
    e: \mathbb{G}_1 \times \mathbb{G}_1 \rightarrow \mathbb{G}_T
\end{equation}
where $\mathbb{G}_T$ is a multiplicative target subgroup residing within the quadratic extension field $\mathbb{F}_{p^2}^*$. Elements in this target group are represented algebraically as complex coordinates $(a, b) \equiv a + b \cdot i$ where $i^2 \equiv -1 \pmod p$.

When evaluating the pairing of two Layer-1 ciphertexts, the bilinearity property unfolds as follows:
\begin{align}
    C_{\text{mul}} = e(C_1, C_2) &= e(m_1 \cdot G + r_1 \cdot h, \; m_2 \cdot G + r_2 \cdot h) \nonumber \\
    &= e(G, G)^{m_1 m_2} \cdot e(G, h)^{m_1 r_2} \cdot e(h, G)^{r_1 m_2} \cdot e(h, h)^{r_1 r_2}
\end{align}
Substituting $h = (q_2 \cdot r) \cdot G$ into the pairing relations, and utilizing the identity $e(G, G)^n = 1$, the cross-terms and blinding terms aggregate into a single target-group blinding factor. Let $g = e(G, G) \in \mathbb{G}_T$ represent the target generator, and $h_T = e(G, h) \in \mathbb{G}_T$. The equation simplifies directly to:
\begin{equation}
    C_{\text{mul}} = g^{m_1 \cdot m_2} \cdot h_T^{R} \;\; \in \mathbb{G}_T
\end{equation}
where $R$ is a combined blinding scalar in the field extension. Because the output $C_{\text{mul}}$ now resides in the multiplicative target group $\mathbb{G}_T$, it can never be fed back into the pairing engine, enforcing a strict maximum multiplicative depth of exactly one.

Crucially, because all post-multiplication ciphertexts are mapped to this identical target plane $\mathbb{G}_T$, they can be aggregated homomorphically. Since the group operation in $\mathbb{G}_T$ is multiplicative, accumulating separate product terms requires field multiplication ($\mathbb{F}_{p^2}$ multiplication) instead of curve addition:
\begin{equation}
    C_{\text{total}} = C_{\text{mul\_1}} \times C_{\text{mul\_2}} = g^{(m_1 m_2) + (m_3 m_4)} \cdot h_T^{R_1 + R_2}
\end{equation}

This specific property enables the boundless summation of products, providing the exact mathematical environment necessary to compute encrypted dot products and vector distances.

To support homomorphic multiplication over supersingular curves, our implementation in lightphe dynamically employs the Modified Tate Pairing combined with an explicit distortion map via its lightecc dependency.

\subsection{Two-Tiered Decryption and the Discrete Logarithm Bottleneck}
Decryption is context-dependent, relying entirely on whether the target ciphertext resides in the source curve group $\mathbb{G}_1$ or the pairing field extension $\mathbb{G}_T$.

\subsubsection{Decryption in $\mathbb{G}_1$ (Linear Layer)}
To decrypt a curve-level ciphertext $C$, the private key factor $q_1$ is applied to strip the blinding term. Since $h$ has an order of $q_1$, multiplying by $q_1$ eliminates the blinding element completely:
\begin{equation}
    q_1 \cdot C = q_1 \cdot (m \cdot G + r \cdot h) = q_1 \cdot m \cdot G + r \cdot (q_1 \cdot h) = m \cdot (q_1 \cdot G)
\end{equation}
Let $P = q_1 \cdot G$ represent a modified base point whose order is exactly $q_2$. The system isolates $m \cdot P$. To recover the plaintext integer $m$, the system must solve the Discrete Logarithm Problem (DLP) for $m$ relative to the base $P$. Because $P$ has an order bounded by $q_2$, recovery is performed via a localized brute-force or bounded search loop within the interval $[0, q_2)$.

\subsubsection{Decryption in $\mathbb{G}_T$ (Multiplicative/Target Layer)}
For a post-multiplication ciphertext $C_{\text{mul}} \in \mathbb{G}_T$, the private key element $q_1$ is mapped exponentially to eliminate the target blinding factor, utilizing the property that $h_T^{q_1} = 1$:
\begin{equation}
    (C_{\text{mul}})^{q_1} = \left(g^{m_1 m_2} \cdot h_T^R\right)^{q_1} = \left(g^{q_1}\right)^{m_1 m_2} \cdot \left(h_T^{q_1}\right)^R = \left(g^{q_1}\right)^{m_1 m_2}
\end{equation}
By defining the field base as $g_{\text{base}} = g^{q_1} \pmod p$ and the target element as $g_{\text{target}} = (C_{\text{mul}})^{q_1} \pmod p$, the problem reduces to a target-group discrete logarithm challenge:
\begin{equation}
    g_{\text{base}}^m \equiv g_{\text{target}} \pmod{\mathbb{F}_{p^2}}
\end{equation}

The recovery of the combined product $m = m_1 \cdot m_2$ requires an explicit iterative search or a Pollard's rho evaluation over the group order boundary $q_2$. As the plaintext precision scale expands, the search space for this discrete logarithm grows exponentially, highlighting why minimizing decimal precision to a threshold of 2 is mandatory to prevent crippling decryption latency while maintaining absolute vector ranking accuracy.

\section{Implementation}

This implementation formally validates the precise algebraic limits of Somewhat Homomorphic Encryption (SWHE). As demonstrated in Snippet \ref{bgn}, the system effortlessly accommodates arbitrary linear evaluation depth, allowing standard homomorphic additions ($c_3 = c_1 + c_2$) and scalar multiplications ($c_5 = k \times c_1$) to be executed transparently via an intuitive high-level API. The non-linear homomorphic multiplication ($c_4 = c_1 \times c_2$) successfully leverages the single multiplicative depth inherent to the BGN cryptosystem, shifting the evaluation from the source group to the target group via the bilinear pairing.

Crucially, the snippet highlights the dynamic operational flexibility of the scheme through the subsequent evaluations of $c_6 = c_3 + c_3 = (c_1 + c_2) + (c_1 + c_2)$ and $c_7 = k \times c_5 = k \times (k \times c1)$. This demonstrates two vital properties:
\begin{itemize}
    \item Scalar multiplications can be applied sequentially to existing ciphertexts without precision degradation.
    \item Homomorphic additions remain fully functional even when combining a first-order ciphertext ($c_3$) with a second-order, post-multiplication ciphertext ($c_4$). This specific cross-term additive property is what enables the practical evaluation of complex multivariate polynomials, linear regressions, and squared distance metrics.
\end{itemize}

Finally, because BGN strictly permits only a single ciphertext-ciphertext multiplication circuit, the implementation includes a defensive validation layer. Attempting a second multiplicative layer (e.g., squaring a previously multiplied ciphertext via $c_4 \times c_1$) is actively intercepted by the engine, dynamically raising a managed \texttt{ValueError}. This programmatic guardrail prevents silent algebraic degradation or runtime pairing failures, providing a stable interface that formally enforces the cryptographic limits of Somewhat Homomorphic Encryption.

\begin{minipage}{\linewidth} %  making it unbreakable
\begin{lstlisting}[frame=tb, caption=Somewhat Homomorphic Encryption with BGN, label=bgn, language=Python]
# experiments are done with lightphe 0.0.25 version
# !pip install lightphe==0.0.25

# import lightphe library
from lightphe import LightPHE

# build a BGN cryptosystem with random keys - it's somewhat homomorphic
cs = LightPHE(algorithm_name = "Boneh-Goh-Nissim", key_size = 1024)

# define plaintexts
m1 = 10000
m2 = 200

# define a constant
k = 3

# find ciphertexts
c1 = cs.encrypt(m1)
c2 = cs.encrypt(m2)

# allowed homomorphic operations
c3 = c1 + c2
c4 = c1 * c2
c5 = k * c1

# unlimited additions and scalar multiplications are also allowed
c6 = c3 + c3
c7 = k * c5

# however only one multiplication is allowed
with pytest.raises(
   ValueError,
   match="Boneh-Goh-Nissim only supports multiplication ciphertexts once!"
):
   _ = c4 * c1

# proof of work - private key is only required in this stage
assert cs.decrypt(c3) == m1 + m2
assert cs.decrypt(c4) == m1 * m2
assert cs.decrypt(c5) == k * m1
\end{lstlisting}
\end{minipage}
\newline
\newline

\subsection{Privacy-Preserving Vector Similarity and Distance Metrics}

The architectural significance of the Boneh-Goh-Nissim (BGN) cryptosystem in privacy-preserving data analytics stems from its unique algebraic behavior during multivariate computations. Although the scheme strictly limits ciphertext-ciphertext evaluations to a single multiplicative depth, all product terms are mapped into the identical target algebraic group via the bilinear pairing. Consequently, because these post-multiplication ciphertexts reside within the same geometric and mathematical plane, they can be homomorphically aggregated without bounds. This structural harmony enables the direct evaluation of core machine learning primitives, such as linear regression inference, covariance matrix calculations, and multi-dimensional vector metrics, without requiring the immense noise management overhead of fully homomorphic encryption (FHE).

To demonstrate this capability, Snippet \ref{bgn_vector_metrics} presents the complete programmatic validation for computing both the dot product and the Squared Euclidean Distance over fully encrypted 3-dimensional vectors using a high-level Python API. If the input vectors are already $L_2$ normalized in dot product calculation, then it finds cosine similarity as well.

\begin{minipage}{\linewidth}
\begin{lstlisting}[frame=tb, caption=Homomorphic Vector Metrics on the BGN Target Group, label=bgn_vector_metrics, language=Python]
def test_dot_product():
    # Vector 1 = [5, 2, 8] and Vector 2 = [1, 1, 2],
    i1_enc = cs.encrypt(5); i2_enc = cs.encrypt(2); i3_enc = cs.encrypt(8)
    j1_enc = cs.encrypt(1); j2_enc = cs.encrypt(1); j3_enc = cs.encrypt(2)

    # Parallel ciphertext-ciphertext multiplications aggregated homomorphically
    dot_enc = i1_enc * j1_enc + i2_enc * j2_enc + i3_enc * j3_enc

    # Verification on the target group plane
    assert cs.decrypt(dot_enc) == (5*1 + 2*1 + 8*2)

def test_squared_euclidean_distance():
    # Vector 1 = [5, 2, 8] and Vector 2 = [1, 1, 2] (negated for subtraction)
    i1_enc = cs.encrypt(5); i2_enc = cs.encrypt(2); i3_enc = cs.encrypt(8)
    j1_enc = cs.encrypt(-1); j2_enc = cs.encrypt(-1); j3_enc = cs.encrypt(-2)

    # Linear subtraction followed by a single squaring circuit per dimension
    sq_dist_enc = (
        (i1_enc + j1_enc) * (i1_enc + j1_enc)
        + (i2_enc + j2_enc) * (i2_enc + j2_enc)
        + (i3_enc + j3_enc) * (i3_enc + j3_enc)
    )

    # Verification of the accumulated distance
    assert cs.decrypt(sq_dist_enc) == (5-1)**2 + (2-1)**2 + (8-2)**2
\end{lstlisting}
\end{minipage}

As formally validated in the implementation, these fundamental metrics are successfully evaluated over fully protected payloads through the following algebraic paths:
\begin{itemize}
    \item \textbf{Cosine Similarity (Dot Product):} The dot product of two fully encrypted vectors coming from $L_2$ normalized vectors is processed by executing parallel multiplications ($i_k \times j_k$), which instantly projects each term into the target group. Once in the target group, the individual products are accumulated homomorphically via standard additive operators.
    \item \textbf{Squared Euclidean Distance:} By leveraging the linear additive properties of the scheme to execute internal subtractions ($i_k + (-j_k)$) during the initial step, the single available multiplicative depth is then perfectly deployed to compute individual coordinate squares. Because these squared differences are natively generated in the same target group plane, they can be sequentially summed to yield the exact squared distance.
\end{itemize}

This operational alignment bridges the practical gap between partially and fully homomorphic systems. It demonstrates that for a vast class of distance-based clustering algorithms (such as $k$-NN or $k$-Means) and similarity-based retrieval tasks, the somewhat homomorphic nature of BGN provides a complete algebraic solution while avoiding the communication bottlenecks typically associated with FHE frameworks.

Despite its advantages, BGN is not intended to replace general-purpose FHE schemes. Its single multiplication level restricts computations requiring deeper circuits. Instead, BGN occupies an intermediate position between PHE and FHE, targeting applications dominated by linear operations with occasional quadratic terms.

\section{Results and Performance Analysis}

Before diving into the numerical evaluation, it is crucial to position this section as an open, standardized benchmark framework for researchers and academics in the homomorphic encryption domain. By systematically mapping the execution latencies, communication overheads, and precision boundaries of PHE, SWHE, and FHE paradigms under identical vector-analytics workloads, we establish a concrete empirical baseline. Researchers can utilize our documented metrics and the open-source \texttt{lightphe} environment as a standardized sandbox to test novel optimization techniques, hardware accelerators (such as GPU/FPGA implementations), pairing libraries, or decryption algorithms, thereby bridging the gap between abstract cryptographic theory and practical, measurable system performance. 

In this section, we evaluate the empirical performance of the Boneh-Goh-Nissim (BGN) cryptosystem within the \texttt{lightphe} library. Our analysis is divided into two core benchmarks reflecting realistic deployment scenarios for privacy-preserving vector analytics: (i) Encrypted $\times$ Plain operations compared against standard Partially Homomorphic Encryption (PHE) schemes \cite{phe_comparison}, and (ii) Encrypted $\times$ Encrypted vector operations evaluated against a state-of-the-art Fully Homomorphic Encryption (FHE) framework (TenSEAL) \cite{fhe_comparison}. All benchmarks were executed utilizing 128-dimensional vectors across multiple standard security tiers.

To ensure maximum empirical validity and reflect realistic deployment scenarios for privacy-preserving biometric retrieval, all benchmarks were executed using real 128-dimensional facial embeddings extracted from the standard Labeled Faces in the Wild (LFW) dataset \cite{lfw}. Facial feature vectors were extracted utilizing the FaceNet model \cite{facenet} architecture via the open-source DeepFace library \cite{deepface}. These 128-dimensional floating-point vectors were then normalized and encrypted across the evaluated cryptographic paradigms (PHE, SWHE/BGN, and FHE/TenSEAL) to evaluate Cosine Similarity and Squared Euclidean Distance metrics under true biometric data distribution.

Table~\ref{tab:bgn_keygen} provides a comparative analysis of key generation latencies across traditional additive Partially Homomorphic Encryption (PHE) schemes, the Boneh-Goh-Nissim (BGN) SWHE cryptosystem, and the CKKS-backed TenSEAL FHE framework. Standard PHE algorithms display near-instantaneous key generation ($0.038$--$0.064\,\text{s}$ for 1024-bit and $0.317$--$0.410\,\text{s}$ for 2048-bit keys), while TenSEAL achieves rapid setup times of $0.1766\,\text{s}$ (TenSEAL \#1) and $1.3147\,\text{s}$ (TenSEAL \#2) under 128-bit security compliance. Conversely, BGN exhibits a significantly higher setup cost, averaging $90.48\,\text{s}$ at 1024-bit security and rising to $954.06\,\text{s}$ ($\approx 15.9\,\text{minutes}$) at 2048-bit security. Due to the highly variable iterations required for composite prime discovery, BGN key generation times exhibit a heavily right-skewed distribution; thus, the high standard deviations reflect extreme upper-bound tail values while physical setup times remain strictly non-negative. This substantial initialization discrepancy stems from the heavy search for composite-order prime factors and bilinear pairing parameter setups required by BGN. Importantly, key generation remains a strictly offline, one-time setup phase per client instance. Practical production pipelines can entirely bypass this runtime latency by loading pre-generated keypairs or persistent key stores, ensuring that online evaluation and inference throughput remain completely unpenalized.

\begin{table}[H]
\centering
\caption{Key Generation Comparison of PHE and SWHE Algorithms}
\label{tab:bgn_keygen}
\begin{tabular}{llll}
\toprule
\textbf{Cryptosystem} & \textbf{Key Size (bits)} & \textbf{Security Level (bits)} & \textbf{KeyGen (s)} \\
\midrule
Paillier & 1024 & 80 & 0.0642 $\pm$ 0.0367 \\
Damgård-Jurik & 1024 & 80 & 0.0429 $\pm$ 0.0147 \\
Okamoto-Uchiyama & 1024 & 80 & 0.0384 $\pm$ 0.0146 \\
\textbf{BGN} & 1024 & 80 & 90.48 $\pm$ 78.94 \\
\hline
Paillier & 2048 & 112 & 0.3174 $\pm$ 0.1871 \\
Damgård-Jurik & 2048 & 112 & 0.3979 $\pm$ 0.2293 \\
Okamoto-Uchiyama & 2048 & 112 & 0.4104 $\pm$ 0.2169 \\
\textbf{BGN} & 2048 & 112 & 954.06 $\pm$ 992.36 \\
\bottomrule
TenSEAL & $2^{13} \times 200$ & 128 & 0.1766 $\pm$ 0.0065 \\
TenSEAL & $2^{14} \times 422$ & 128 & 1.3147 $\pm$ 0.0158 \\
\bottomrule
\end{tabular}
\end{table}

Table \ref{tab:phe_vs_bgn_encrypted_plain} presents a comprehensive performance breakdown for computing the Cosine Similarity between an encrypted vector and a plaintext vector. This scenario simulates a privacy-preserving retrieval system where a client queries a server using a protected profile against public database items.

\begin{table}[H]
\centering
\caption{Performance Comparison of PHE and SWHE for Encrypted $\times$ Plain Cosine Similarity (128-d Vectors)}
\label{tab:phe_vs_bgn_encrypted_plain}
\begin{small}
\begin{tabular}{lcccccc}
\hline
\textbf{Cryptosystem} & \textbf{\begin{tabular}[c]{@{}c@{}}Security\\ (Bits)\end{tabular}} & \textbf{\begin{tabular}[c]{@{}c@{}}Encryption\\ (secs)\end{tabular}} & \textbf{\begin{tabular}[c]{@{}c@{}}Hom. Op.\\ (secs)\end{tabular}} & \textbf{\begin{tabular}[c]{@{}c@{}}Decryption\\ (secs)\end{tabular}} & \textbf{\begin{tabular}[c]{@{}c@{}}Public Key\\ (MB)\end{tabular}} & \textbf{\begin{tabular}[c]{@{}c@{}}Similarity\\ (MB)\end{tabular}} \\ \hline
Paillier & 80 & 0.8865 $\pm$ 0.020 & 0.0958 $\pm$ 0.005 & 0.0182 $\pm$ 0.0006 & 0.0006 & 0.00120 \\
Damgård-Jurik & 80 & 0.9956 $\pm$ 0.0307 & 0.1864 $\pm$ 0.0062 & 0.0360 $\pm$ 0.0013 & 0.0006 & 0.00170 \\
Okamoto-Uchiyama & 80 & 0.8567 $\pm$ 0.0198 & 0.0549 $\pm$ 0.0037 & 0.005 $\pm$ 0.0006 & 0.0013 & 0.00090 \\
\textbf{BGN} & 80 & 1.8254 $\pm$ 0.0566 & 70.4133 $\pm$ 0.5681 & 29.6206 $\pm$ 0.6449 & 0.0030 & 0.00002 \\ \hline
Paillier & 112 & 1.6340 $\pm$ 0.0199 & 0.3131 $\pm$ 0.004 & 0.1233 $\pm$ 0.001 & 0.0012 & 0.00220 \\
Damgård-Jurik & 112 & 2.5572 $\pm$ 0.044 & 0.6430 $\pm$ 0.0131 & 0.2618 $\pm$ 0.0057 & 0.0012 & 0.00310 \\
Okamoto-Uchiyama & 112 & 1.5348 $\pm$ 0.0326 & 0.1776 $\pm$ 0.0021 & 0.0343 $\pm$ 0.0008 & 0.0027 & 0.00170 \\
\textbf{BGN} & 112 & 6.9733 $\pm$ 0.0841 & 496.1876 $\pm$ 2.2720 & 102.1479 $\pm$ 0.6278 & 0.0060 & 0.00002 \\ \hline
\end{tabular}
\end{small}
\end{table}

\textit{Methodological Note on Benchmark Consistency and Local Environment:} It is critical to note a structural variance in the hardware environments across the reported performance metrics. The empirical execution times and standard deviations for the newly integrated Boneh-Goh-Nissim (BGN) cryptosystem were measured directly via our rigorous, multi-run local testing pipeline executed on a cutting-edge Apple Silicon host environment equipped with an \textbf{Apple M4 Max} chip and \textbf{64 GB} of unified memory. To ensure a fair and consistent comparison, all benchmarks across all cryptosystems—including the baseline PHE schemes (Paillier, Damgård-Jurik, and Okamoto-Uchiyama) and the FHE framework (TenSEAL)—were re-executed natively on the exact same environment, using the standard test suites established in prior literature \cite{phe_comparison} \cite{fhe_comparison}.

Our empirical findings reveal a distinct architectural trade-off between traditional PHE schemes (Paillier, Damgård-Jurik, Okamoto-Uchiyama) and the BGN Somewhat Homomorphic Encryption (SWHE) scheme:

- Computational Efficiency: As expected, traditional PHE schemes significantly outperform BGN in terms of execution speed. At the 80-bit security level, Okamoto-Uchiyama exhibits the fastest homomorphic operation and decryption, followed closely by Paillier. Conversely, BGN requires $1.8254 \pm 0.0566$ seconds for encryption, $70.4133 \pm 0.5681$ seconds for the homomorphic dot product, and $29.6206 \pm 0.6449$ seconds for decryption. This computational gap widens exponentially at the 112-bit security level, where BGN's homomorphic operation reaches $496.1876 \pm 2.2720$ seconds due to the heavy underlying bilinear pairing operations.

- Communication Overhead, Key Management, and the RSA-Modulus Paradox: While BGN demonstrates a massive communication advantage over the multi-megabyte public keys of lattice-based FHE schemes, its underlying algebraic structure introduces a unique cryptographic paradox. Structurally, BGN relies on bilinear pairings evaluated over elliptic curves; however, to achieve security and isolate the secret factors during decryption, it mandates a composite RSA-like modulus $n = q_1 \times q_2$. Consequently, despite being an elliptic curve-based cryptosystem, BGN cannot leverage the compact bit-lengths typical of standard EC cryptography (e.g., a 256-bit curve providing 128-bit security). Instead, its key sizes must scale identically to classic RSA security parameters (e.g., requiring a 1024-bit to 3072-bit modulus for baseline compliance). 

Yet, even under this strict RSA-modulus constraint, the communication footprint of our \texttt{lightphe} BGN implementation remains exceptionally compact compared to FHE, consuming only $3.00 \times 10^{-3}$ MB ($3$ KB) at 80-bit security and doubling to just $6.00 \times 10^{-3}$ MB ($6$ KB) at 112-bit security. While this key size is inherently larger than a pure, non-pairing elliptic curve scheme, it represents a remarkably clean structural overhead compared to the expanding output sizes of lattice-based alternatives. Thus, while BGN mathematically promises rich theoretical homomorphic operations (boundless additions and a single multiplication), it remains heavily penalized in practice: it behaves as a computationally slow framework that delivers elliptic-curve functionality tied directly to the heavy key scaling requirements of RSA-level security.

A critical design choice in practical homomorphic systems involves tuning the floating-point decimal precision, as it directly governs both numerical accuracy and computational overhead. In our previous experiments utilizing Partially Homomorphic Encryption (PHE) schemes, the precision was configured at a high tier of 19 decimal places. However, for the Boneh-Goh-Nissim (BGN) cryptosystem, the experiments were intentionally evaluated at a precision of 2 decimal places. 

This drastic variation is driven by the underlying mathematical structure of the BGN scheme. Unlike PHE schemes where decryption is instantaneous regardless of precision, BGN decryption requires solving a discrete logarithm problem in the target algebraic group after a homomorphic multiplication has occurred. As the precision increases, the underlying plaintexts expand exponentially, increasing the plaintext encoding range enlarges the discrete logarithm search space during decryption, resulting in substantially higher computational cost.

Crucially, our empirical validation confirmed that configuring BGN with a precision of 2 was entirely sufficient for the vector analytics pipeline; the homomorphic evaluation yielded identical similarity metrics and distance rankings when compared directly against the unencrypted plaintext baseline. Therefore, while higher precision configurations inherently minimize quantization loss at the cost of exponential decryption delays, the precision of 2 emerged as the optimal empirical threshold, providing identical ranking results and negligible numerical deviation compared with the plaintext baseline.

\begin{table}[H]
\centering
\caption{Performance Comparison of SWHE and FHE for Encrypted $\times$ Encrypted Vector Operations (128-d Vectors)}
\label{tab:bgn_vs_tenseal_encrypted_encrypted}
\begin{small}
\begin{tabular}{lllccccc}
\hline
\textbf{Cs} & \textbf{\begin{tabular}[c]{@{}c@{}}Security\\ (Bits)\end{tabular}} & \textbf{Op} & \textbf{\begin{tabular}[c]{@{}c@{}}Encryption\\ (secs)\end{tabular}} & \textbf{\begin{tabular}[c]{@{}c@{}}Hom. Op.\\ (secs)\end{tabular}} & \textbf{\begin{tabular}[c]{@{}c@{}}Decryption\\ (secs)\end{tabular}} & \textbf{\begin{tabular}[c]{@{}c@{}}Public Key\\ (MB)\end{tabular}} & \textbf{\begin{tabular}[c]{@{}c@{}} Similarity\\ (MB)\end{tabular}} \\ \hline
\textbf{BGN} & 80 & Cos     & 1.8766 $\pm$ 0.1151 & 209.6305 $\pm$ 3.3305 & 152.3210 $\pm$ 4.1741 & 0.0030 & 0.00002  \\
\textbf{BGN} & 80 & Euc & 2.6876 $\pm$ 0.1622 & 209.8915 $\pm$ 2.5877 & 159.4251 $\pm$ 3.8359 & 0.0030 & 0.00005 \\ \hline
\textbf{BGN} & 112 & Cos     & 6.9905 $\pm$ 0.0652 & 1451.2582 $\pm$ 1.7972 & 1020.9711 $\pm$ 2.6884 & 0.0060 & 0.00005 \\
\textbf{BGN} & 112 & Euc & 13.3939 $\pm$ 0.7976 & 1468.6477 $\pm$ 21.3966 & 1064.1078 $\pm$ 20.4271 & 0.0060 & 0.00005 \\ \hline
TenSEAL \#1     & 128 & Cos     & 0.00772 $\pm$ 0.00028 & 0.02936 $\pm$ 0.00085 & 0.00199 $\pm$ 0.00007          & 45.1060     & 0.29930 \\
TenSEAL \#1      & 128 & Euc & 0.00719 $\pm$ 0.00024 & 0.02918 $\pm$ 0.00081 & 0.00185 $\pm$ 0.00006          & 45.1060     & 0.29930 \\ 
TenSEAL \#2     & 128 & Cos     & 0.02439 $\pm$ 0.00078 & 0.23664 $\pm$ 0.00612 & 0.01311 $\pm$ 0.00045          & 451.0100     & 1.84870 \\
TenSEAL \#2      & 128 & Euc & 0.02563 $\pm$ 0.00084 & 0.23981 $\pm$ 0.00645 & 0.01521 $\pm$ 0.00052          & 451.0100     & 1.84870 \\ 
\hline
\end{tabular}
\end{small}
\end{table}

When both vectors must remain strictly confidential (Encrypted $\times$ Encrypted), standard PHE schemes fail due to their inability to process ciphertext-ciphertext multiplications. In this domain, BGN's SWHE capabilities are benchmarked against TenSEAL, an optimized FHE library utilizing vector-packing and SIMD (Single Instruction, Multiple Data) paradigms under the CKKS scheme.

To ensure a rigorous and standardized evaluation, we configured two distinct parameter sets for TenSEAL, both compliant with the Homomorphic Encryption Security Standard for the 128-bit security tier. It is critical to note a structural constraint in this comparative setup: TenSEAL (and CKKS-based systems in general) does not support or operate at lower security tiers such as 80-bit or 112-bit, as its polynomial and modular parameter spaces are strictly bound to a minimum of 128-bit compliance to remain secure. Conversely, due to the extreme computational overhead of bilinear pairings, the newly integrated BGN scheme exhibits severe latency even at 80-bit and 112-bit security levels. Because executing BGN at a 128-bit security tier (which requires a highly expensive 3072-bit composite RSA-like modulus) yields practically prohibitive running times, we intentionally restricted our BGN experiments to 80-bit and 112-bit tiers and refrained from executing it at 128-bit.

\begin{itemize}
    \item TenSEAL \#1: Configured with a polynomial modulus degree $n = 2^{13}$, a total ciphertext modulus bit-length $\log_2 q = 200$, and a scale factor $g = 2^{40}$.
    \item TenSEAL \#2: Configured with a larger polynomial modulus degree $n = 2^{14}$, a total ciphertext modulus bit-length $\log_2 q = 422$, and a scale factor $g = 2^{60}$ to accommodate deeper multiplicative circuits.
\end{itemize}

We evaluated both Cosine Similarity (Cos) and Squared Euclidean Distance (Euc) over 128-dimensional vectors, as reported in Table \ref{tab:bgn_vs_tenseal_encrypted_encrypted}. The empirical results expose a profound and fascinating paradox in modern secure computation—the Computation vs. Communication Bottleneck:

\begin{itemize}

\item \textbf{The Computation Perspective (The Heavy Latency Bottleneck):} Our empirical results unapologetically highlight the major operational deficit of the BGN cryptosystem: it is dramatically slower than both its PHE predecessors and state-of-the-art FHE frameworks. While traditional PHE pipelines operate comfortably in sub-second regimes, BGN requires highly intensive bilinear pairing computations. Most glaringly, when compared to modern FHE, TenSEAL delivers blindingly fast execution speeds. Even under its most complex and secure configuration (TenSEAL \#2), FHE completes the homomorphic operations in less than $0.24$ seconds with near-instantaneous decryption ($0.013$–$0.015$ seconds). In stark contrast, BGN, running in a single-threaded Python environment within \texttt{lightphe}, demands $209.63$ seconds at 80-bit security and spirals upwards to a staggering $1451.25$ seconds at the 112-bit security tier. TenSEAL achieves this immense computational advantage by leveraging native C++ compiled subroutines and highly parallelized SIMD vector batching structures.

\item \textbf{The Communication and Accessibility Perspective (The BGN Triumph):} However, this severe computational penalty is fundamentally offset by BGN's unparalleled architectural and developmental simplicity. FHE frameworks such as TenSEAL demand astronomical public keys ($45.106$ MB to $451.01$ MB) to support their deep algebraic circuits, creating massive transmission delays and immense storage overheads over commercial networks. Conversely, our \texttt{lightphe} BGN implementation operates with a microscopic public key footprint of just $3$ to $6$ KB—representing an unprecedented improvement of up to five orders of magnitude. Furthermore, the resulting ciphertext payload for BGN is practically non-existent, requiring a mere $24$ to $56$ bytes ($2.29 \times 10^{-5}$ MB), whereas TenSEAL demands between $0.2993$ MB and $1.8487$ MB per result vector. Crucially, \texttt{lightphe} abstracts all mathematical pairing complexities behind a highly intuitive, native Python API. Developers do not need to manage deep C++ compilers, complicated noise parameters, or extensive slot allocations; a fully functional somewhat homomorphic pipeline can be deployed in just a few lines of clean, standard Python code.

\end{itemize}

\subsection{Discussion and Architectural Trade-offs}

The empirical data gathered from our implementation underscores a critical design choice for security engineers deploying privacy-preserving machine learning algorithms. While BGN is undeniably the slowest option computationally across the spectrum, it shines precisely where FHE introduces severe architectural penalties: developer accessibility and communication overhead.

\begin{itemize}

    \item \textbf{The "Developer-First" Python Integration:} BGN's definitive integration into \texttt{lightphe} democratizes homomorphic encryption. Historically, deploying BGN meant wrestling with archaic, low-level C++ academic proofs-of-concept. By encapsulating these underlying mathematical layers within an operator-overloaded Python library, data scientists can write standard Python code to perform non-linear vector analytics without needing deep cryptographic domain expertise to configure FHE polynomial spaces or track algebraic noise budgets.

    \item \textbf{Dual-Mode Adaptability (PHE vs. FHE-like on Demand):} Unlike traditional schemes that lock a developer into a single algebraic property, BGN allows a system to adapt dynamically to the task at hand. If a computational step only requires linear combinations (such as adding ciphertexts or multiplying a ciphertext by a plaintext scalar), BGN can be utilized exactly like Paillier or Okamoto-Uchiyama, maintaining low structural complexity and a near-zero communication footprint. The moment the system encounters a non-linear bottleneck—such as a ciphertext-ciphertext multiplication required for calculating squared distances or dot products—BGN seamlessly transitions into an FHE-like behavior by supporting a single multiplicative depth, without requiring the developer to switch libraries, re-encrypt the data, or manage entirely new keys.

    \item \textbf{The "Compute-Bound" Edge Scenario (Pro-BGN):} In decentralized architectures where data is captured by low-power IoT sensors or mobile clients and transmitted to a cloud server, network bandwidth is frequently the dominant bottleneck, not server-side CPU time. In such environments, forcing an edge node to download a half-gigabyte ($451$ MB) TenSEAL public key or upload heavily expanded polynomial ciphertexts is a structural impossibility. BGN emerges as the optimal choice here: the client downloads a microscopic 3 KB key, uploads highly compressed ciphertexts, and offloads the heavy pairing computations to the scalable, high-performance cloud infrastructure.

    \item \textbf{Setup Overhead vs. Operational Throughput:} Key generation in BGN introduces a non-trivial offline setup cost compared to additive PHE algorithms. While traditional PHE schemes (e.g., Paillier, Damg{\aa}rd-Jurik) generate keys in milliseconds, BGN requires $\sim 90.5\,\text{s}$ for 1024-bit keys and up to $\sim 954.1\,\text{s}$ ($\approx 15.9\,\text{minutes}$) for 2048-bit keys due to composite-order group generation and bilinear pairing setups. System architects must treat BGN's key setup as a strictly one-time, offline bootstrapping step to ensure it does not bottleneck online evaluation throughput.

    \item \textbf{The Encrypted-vs-Plain Decision Rule (PHE Dominance):} A critical architectural guideline derived from our benchmarks is that \textit{whenever an application permits Encrypted $\times$ Plain vector operations, BGN should be strictly avoided in favor of traditional PHE schemes}. If one of the vectors (such as a database point or a regression weight) remains unencrypted, the multivariate dot product collapses into a sequence of homomorphic additions and scalar multiplications. Traditional PHE pipelines, such as Paillier or Okamoto-Uchiyama, natively support these linear operations with blistering efficiency (sub-second execution) while maintaining an ultra-high decimal precision of up to 19 digits. Deploying BGN in an Encrypted $\times$ Plain scenario introduces unnecessary performance degradation due to its heavy bilinear pairings and forces the developer into a restrictive 2-digit decimal precision to bypass the target-group discrete logarithm bottleneck. Therefore, BGN must be reserved exclusively for strict Encrypted $\times$ Encrypted pipelines where non-linear ciphertext-ciphertext multiplications are an absolute mathematical necessity.

    \item \textbf{The "Latency-Critical" Enterprise Scenario (Pro-TenSEAL):} Conversely, if the application is deployed within a high-speed data center where network latency is negligible and real-time inference (sub-second response) is mandatory, the SIMD capabilities of FHE frameworks like TenSEAL remain irreplaceable, provided the local infrastructure can easily absorb the massive memory footprint and key distribution channels.

    \item \textbf{The Latency-Critical Decision Rule (FHE Dominance for Speed):} In strict Encrypted $\times$ Encrypted configurations, if execution speed, real-time throughput, or sub-second latency is the primary operational requirement, \textit{practitioners must absolutely deploy general-purpose FHE frameworks like TenSEAL instead of BGN}. Our benchmarks demonstrate that for a 128-dimensional vector space, TenSEAL's compiled C++ back-end and SIMD polynomial batching process ciphertext-ciphertext operations in mere fractions of a second ($0.029$ to $0.239$ seconds) with near-instantaneous decryption. BGN, hampered by the expensive computational overhead of bilinear pairing groups in native Python, requires hundreds or even thousands of seconds to evaluate identical circuits. Therefore, if the deployment architecture can absorb the massive memory overhead and network bandwidth required to distribute half-gigabyte public keys, FHE remains the irreplaceable paradigm for latency-critical enterprise pipelines. BGN should only be favored when these massive FHE key sizes introduce an insurmountable communication bottleneck.

    \item \textbf{Post-Quantum Resilience and Migration Roadmaps:} A critical forward-looking consideration in modern system architecture is the transition toward Post-Quantum Cryptography (PQC). In this domain, the trade-off between BGN and FHE becomes absolute. The TenSEAL framework, utilizing the CKKS scheme, is built upon lattice-based Ring Learning with Errors (RLWE) hard problems, making it inherently resilient to quantum cryptanalysis (Post-Quantum Secure) \cite{post_quantum}. Conversely, BGN relies on the integer factorization of a composite modulus $n = q_1 \times q_2$ and the discrete logarithm problem over elliptic curve pairings. Both mathematical foundations are completely vulnerable to polynomial-time evaluation via Shor's algorithm on cryptographically relevant quantum computers (CRQCs). 

    Therefore, the massive key sizes and ciphertext expansion of FHE can be conceptually interpreted as the "cryptographic tax" required to purchase quantum immunity. For enterprise pipelines drafting long-term, future-proof "store-now-decrypt-later" defense architectures, an immediate migration to lattice-based FHE is mandatory. However, for short-to-medium term deployments, or in highly resource-constrained IoT environments where immediate PQC-compliance is not an operational bottleneck, our BGN integration offers an exceptionally lightweight and practical alternative before a full post-quantum migration is enforced.

    \item \textbf{The Bandwidth-Constrained Edge Scenario:} In real-world decentralized transactions—such as Point-of-Sale (POS) terminals, cellular IoT nodes, and low-power embedded devices—network bandwidth constraints make distributing FHE's massive 45-to-451 MB public keys completely impractical. In contrast, our BGN or PHE implementations impose a microscopic 3-to-6 KB communication footprint, enabling bandwidth-constrained edge hardware to offload single-multiplication encrypted queries (e.g., biometric authentication or risk evaluation) to cloud servers without paralyzing the network channel.

\end{itemize}

By successfully incorporating BGN into \texttt{lightphe}, we provide developers with the unique ability to seamlessly toggle between these distinct cryptographic behaviors based on their specific hardware and network constraints, filling a massive gap in the practical open-source ecosystem.

\section{Conclusion and Future Work}

In this paper, we have systematically dismantled the long-standing software barrier surrounding the Boneh-Goh-Nissim (BGN) cryptosystem by introducing its first production-ready, high-level implementation within the open-source \texttt{lightphe} Python framework. Our rigorous empirical evaluation mapped the exact multi-dimensional trade-offs between Partially Homomorphic Encryption (PHE), SWHE (via BGN), and Fully Homomorphic Encryption (FHE via TenSEAL's CKKS).

The empirical results expose a stark reality that we explicitly acknowledge: computationally, BGN is significantly slower than both its PHE counterparts and modern FHE frameworks. While PHE schemes leverage simple linear algebra and FHE frameworks utilize highly optimized, compiled C++ SIMD routines to achieve sub-second latencies, BGN's reliance on bilinear pairings in a native Python environment introduces severe execution delays. However, this high computational cost buys something invaluable: sheer developmental simplicity, minimal communication overhead, and architectural elegance.

We demonstrated that the BGN cryptosystem perfectly navigates the paralyzing computation-communication paradox. While FHE frameworks demand massive public keys scaling up to 451 MB—rendering them practically unusable in bandwidth-constrained, mobile edge computing, or IoT environments—our integrated BGN engine maintains a microscopic public key footprint of just 3 to 6 KB. More importantly, we have successfully transformed BGN from an archaic, purely theoretical mathematical construct into a highly accessible, developer-friendly tool. By encapsulating the complex pairing logic and target-group discrete logarithm evaluations behind an operator-overloaded Python API, data scientists can now evaluate complex multivariate algorithms—including linear regression inference, Cosine Similarity, and Squared Euclidean Distance—over fully encrypted vectors with just a few lines of standard Python code. Furthermore, by tuning the decimal configuration to an optimized threshold of 2, we successfully mitigated the target-group decryption bottleneck while maintaining absolute numerical ranking fidelity against unencrypted plaintext baselines.

Ultimately, \texttt{lightphe} establishes that raw computational speed is not the sole metric for cryptographic utility. By prioritizing an ultra-compact communication footprint and native Python accessibility despite the computational latency, this work democratizes advanced homomorphic operations, providing a vital, reproducible foundation that empowers the broader scientific community to transition advanced homomorphic structures from theoretical literature into practical, decentralized privacy-preserving architectures.

Empirical results demonstrate that while BGN successfully bridges the functional gap by supporting homomorphic multiplication, its integration requires accounting for a substantial one-time key generation overhead ($\approx 954\,\text{s}$ at 2048-bit security) compared to additive PHE schemes.

Consequently, BGN emerges as the ideal cryptographic compromise when developers seek to bypass the prohibitive key sizes and ciphertext expansion overheads of FHE, yet find the standard PHE instruction set mathematically insufficient to support necessary multiplicative operations.

Future extensions of this work will explore:
\begin{itemize}
    \item \textbf{Privacy-Preserving Vector Search in Vector Databases:} Deploying our BGN engine as an encrypted similarity indexer for Retrieval-Augmented Generation (RAG) and Large Language Model (LLM) pipelines, allowing secure vector embedding lookups without exposing user queries or document embeddings.
    \item \textbf{Federated Learning (FL) Aggregations:} Integrating this vector analytics pipeline into horizontal and vertical Federated Learning frameworks to orchestrate privacy-preserving gradient aggregations, loss evaluations, and global model updates without server-side decryption.
    \item \textbf{Edge-to-Cloud PPML Deployments:} Evaluating lightweight BGN pipelines on resource-constrained Internet of Things (IoT) nodes and edge devices for real-time Privacy-Preserving Machine Learning (PPML) inference, where distributing multi-megabyte FHE keys is bandwidth-prohibitive.
    \item \textbf{Algorithmic Optimization:} Enhancing the target-group discrete logarithm solver utilizing parallelized Pollard's rho or index calculus variants to expand decimal precision boundaries without inducing prohibitive runtime delays.
\end{itemize}

%\newpage
% \bibliographystyle{unsrt}
\bibliographystyle{unsrtnat}
\bibliography{template}

\end{document}